\documentstyle[12pt,aasms4]{article}

\received{Sept. 23, 1998}
\accepted{Jan. 21, 1999}

\slugcomment{to appear in {\it The Astrophysical Journal}}

\lefthead{Kraemer, Turner, Crenshaw, \& George}
\righthead{The Effects of Intrinsic UV Absorption on the Ionizing Continuum and 
Narrow Emission Line Ratios in Seyfert Galaxies}

\begin{document}

\title{The Effect of Intrinsic UV Absorbers on the Ionizing Continuum and 
Narrow Emission Line Ratios in Seyfert Galaxies}

\author{S. B. Kraemer\altaffilmark{1,4},
T. J. Turner\altaffilmark{2,3},
D. M. Crenshaw\altaffilmark{1},
\& I. M. George\altaffilmark{2}}

\altaffiltext{1}{Catholic University of America,
NASA/Goddard Space Flight Center, Code 681,
Greenbelt, MD  20771.}

\altaffiltext{2}{Universities Space Research Association,
NASA/Goddard Space Flight Center, Code 660,
Greenbelt, MD  20771.}

\altaffiltext{3}{Present address, University of Maryland, 
Baltimore County}

\altaffiltext{4}{Email: stiskraemer@yancey.gsfc.nasa.gov.}

\begin{abstract}

 We explore the effects of UV absorbing material on the shape of the
 EUV continuum radiation emitted by the active galactic nucleus, and
 on the
 relative strengths of emission lines, formed in the narrow line regions of 
 Seyfert galaxies, excited by this continuum. Within a sample of Seyfert 1.5 
 galaxies, objects with 
 flatter soft X-ray slopes tend to have lower values of 
 He~II $\lambda$4686/H$\beta$, which implies a correlation between the 
 observed spectral energy distribution of the ionizing continuum and the 
 narrow emission line strengths. Objects with the flattest soft 
 X-ray continua tend to possess high column density UV absorption and it is 
 plausible that the differences in narrow emission line ratios
 among these galaxies are an indication of the effects of 
 absorbing material internal to the narrow line region, rather than intrinsic 
 differences in continuum shape. We have generated a set of photoionization 
 models to examine the effect of a range of UV absorbers on the 
 ionizing continuum and, hence, the resulting conditions in a 
 typical narrow line cloud. Our results indicate that a 
 low ionization UV 
 absorber with large covering factor will indeed produce the combination of 
 narrow line 
 ratios and soft X-ray spectral characteristics observed in several Seyfert 1.5 
 galaxies. Our results also suggest that low ionization UV absorption may be more 
 common than currently believed.

\end{abstract}


\section{Introduction}

  Since the launch of {\it IUE} in 1978, it has been known that the UV spectra
of a few Seyfert galaxies show absorption lines that are thought to be
intrinsic to the nucleus, as evidenced by their large radial velocities
relative to the host galaxy, large widths, and variability. Although {\it IUE}
studies revealed few Seyfert 1 galaxies that showed intrinsic absorption
(Ulrich 1988), an examination of {\it HST} spectra,
with better sensitivity and/or resolution, 
reveals that more than half ($\sim$ 60\%) of these galaxies showed such
absorption (Crenshaw et al. 1998). Also, approximately half of a sample of Seyfert 1
galaxies observed in X-rays show the presence of an X-ray (``warm'') absorber, characterized by
absorption edges of O~VII and O~VIII (Reynolds 1997; George et al. 1998a).
Although there have been suggestions that the UV and X-ray absorbers
are physically linked (cf. Mathur, Elvis, \& Wilkes 1995; George et al. 1998a), it is also possible that the
absorbing material is comprised of different components at different
radial distances from the central source. In any case, it is clear
that the absorbing material is an important physical component in the
inner regions of Seyfert galaxies and is likely to have a global covering factor 
between 0.5 and 1.0 (Crenshaw et al. 1998).

   It follows that if the covering factor of the absorbing material 
is large, or, at least, co-planar with the narrow-line region (NLR), its
 presence will alter the spectral energy distribution (SED)
of the ionizing continuum to which the NLR gas is exposed.
The effect of the absorption on the
ionizing continuum depends on the ionization state of the absorber
since the conditions in the NLR gas will depend principally
on the number of photons between 13.6 and 100 eV, where the cross-sections
of H, He~I and He~II are greatest. A 
highly ionized X-ray absorber may be transparent at energies below
the O~VII edge ($\sim$ 740 eV), unless it is Compton thick. In such a case, the
NLR will be illuminated by an SED that is effectively intrinsic. 
On the other hand, a UV absorber with
sufficient column density to produce absorption lines of 
Ly$\alpha$, N~V $\lambda$ 1240, and C~IV $\lambda$ 1550 may also
produce a significant edge at the He~II Lyman limit. 
In this case, the effect on the conditions
in the NLR may be significant, and will be revealed by the ratios
of emission lines, particularly He~II $\lambda$4686/H$\beta$ which is
especially sensitive to the SED of the ionizing continuum.
If absorption from lower ionization states, in particular 
Mg~II $\lambda$2800, is present, there may be significant absorption of 
continuum photons near the hydrogen Lyman limit as well. As the attenuation
below 100 eV becomes greater, the average ionization state of the NLR
will drop, while at the same time ionization and heating by X-rays
may dominate, resulting in zones of warm (T $\sim$ 10$^{4}$K), partially
ionized gas.

  In order to examine the possible effects of a low ionization UV absorber 
on the ionizing SED and NLR, we have generated a set of photoionization
models. The photoionization code used in this study was developed by Kraemer 
(1985) and has been described in detail in several previous papers (cf. 
Kraemer et al. 1994). This code has a high energy cutoff at 5 keV. This is 
a sufficiently high cutoff energy to model the effects in the low ionization 
gas, although may be inadequate for an X-ray absorber model.
The modeling methodology and results are discussed in 
detail in the following sections.

\section{The Models}

\subsection{Setting up the Model}

  The purpose of this phase of the modeling is to determine how the modification
of the SED varies for different column densities of the principle UV
absorbers. While the physical conditions within the absorbing gas
are of interest, their examination is not the principle part of this
study. Therefore, we have taken a very simple approach,
and held most parameters fixed while generating the absorber models. 

  The concept that absorption by intervening gas can
modify the ionizing continuum to which the NLR gas is exposed is far from
new. The most detailed treatment has been that of Ferland
\& Mushotzsky (1982), in which they modeled the NLR of NGC 4151,
assuming it was ionized by a continuum modified by a ``leaky absorber''. In 
this model, broad line region (BLR) 
clouds, effectively opaque to the ionizing continuum at EUV energies, cover
90\% of the source, while the remaining 10\% of the ionizing continuum escapes
unattenuated. However, their absorber models predicted ionic column densities
several orders of magnitude larger than those calculated from recent
observations of NGC 4151 (cf. Weymann et al. 1997), and the details of X-ray absorption
were not well known at the time the paper was written. Therefore, it
seems worthwhile to revisit this approach, with the benefit of
better constraints on the absorbers.

  In order to generate these models, we had to assume an intrinsic SED
for the ionizing continuum. Unfortunately, absorption by galactic neutral hydrogen 
makes it all but impossible to get a measurement of the SED in the EUV for 
the vast majority of
AGN, and, thus, there is no direct way to determine the intrinsic shape of
the ionizing continuum.
Although there has been significant effort 
directed towards understanding the shape of the ionizing continuum in active galactic nuclei (AGN), no consensus
has been reached. Recent work by Zheng et al. (1997) and Laor et al. (1997)
suggest that, for QSOs at intermediate redshift, the ionizing continuum 
from the Lyman limit to soft X-ray energies may be characterized by a power 
law, with an index $\alpha$ $\approx$ $-$2. While some lower luminosity AGN,
specifically Seyfert galaxies, are similar, others may have somewhat
flatter indices ($\alpha$ $\approx$ $-$1.5; Korista, Ferland, \& Baldwin 1997).
Korista et al. (1997) have shown that a spectral energy distribution (SED) 
similar to the composite QSO spectrum proposed by Zheng et al. (1995) does not 
possess sufficient He~II ionizing photons to produce the observed equivalent width of the broad 
He~II~$\lambda$ 1640 emission lines in the Seyfert 1 galaxy Mrk 335
(specifically) and, perhaps, AGN in general. Mathews \& Ferland (1987) have 
proposed that the bulk of the necessary He~II ionizing photons in AGN
could arise from the so-called ``big blue bump'' (BBB) (cf. Malkan 
1983).  Therefore, we have generated two sets of models.
For the first, which we will call the ``power-law'' SED, we assumed a broken power law, F$_{\nu}$ $=$ K$\nu^{\alpha}$
where:

\begin{equation}
    \alpha = -1.5,~ 13.6eV \leq h\nu < 1000eV
\end{equation}
\begin{equation}
    \alpha = -0.7, ~h\nu \geq 1000eV. 
\end{equation}

For the other SED, which we will call the ``BBB'' SED, we adopted the parameterization of the Mathews \& Ferland
continuum given in Laor et al. (1997), where 

\begin{equation}
    F_{\nu} = K\nu^{\alpha_{0}}e^{h\nu/KT_{cut}}
\end{equation}
where, $\alpha_{0}$ $=$ $-$0.3 and T$_{cut}$ $=$ 5.4 x 10$^{5}$K. At energies
$\geq$ 200 eV,  F$_{\nu}$ $=$ K$\nu^{\alpha}$, with

 \begin{equation}
    \alpha = -2,~ 200eV \leq h\nu < 1000eV
 \end{equation}

\begin{equation}
    \alpha = -0.7, ~h\nu \geq 1000eV.
\end{equation}

Although it is certainly likely that there is greater variation
in the intrinsic SED among Seyfert galaxies, one can obtain a qualitative 
understanding of how conditions would vary with other power laws and/or 
differing BBB contributions from these results.

  There have been several attempts to determine the density of the gas in which 
the UV resonance line absorption arises. Density estimates based on 
recombination timescale arguments are typically $\geq$ 10$^{5}$ cm$^{-3}$
(cf. Voit et al. 1987; Shull \& Sachs 1993). The presence of weak absorption
from excited states in the spectrum of NGC 4151 would require densities
$\geq$ 10$^{8.5}$ cm$^{-3}$ (Bromage et al. 1985). 
Nevertheless, as often noted (cf. Shields \& Hamann 1997), the ionic column
densities predicted by photoionization models are not particularly sensitive
to density. 
With this in mind, we have assumed a numerical density of atomic 
hydrogen of 1 x 10$^{7}$cm$^{-3}$ typical of the inner NLR in Seyfert
galaxies (cf. Kraemer et al. 1998a), and located the absorbers in the 
intermediate zone between the inner NLR and outer BLR, as suggested by
Espey et al. (1998).  We have assumed solar 
abundances (cf. Grevesse \& Anders 1989) for the UV absorber models, with numerical abundances relative to
hydrogen as follows: He=0.1, C=3.4x10$^{-4}$, O=6.8x10$^{-4}$, 
N=1.2x10$^{-4}$, Ne=1.1x10$^{-4}$, S=1.5x10$^{-5}$, Si=3.1x10$^{-5}$, 
Mg=3.3x10$^{-5}$, Fe=4.0x10$^{-5}$. Cosmic dust was not included.

  In order to examine a range of conditions, we varied the ionization 
parameter for the absorber, U$_{abs}$, where:    

\begin{equation}
U_{abs} = \int^{\infty}_{\nu_0} ~\frac{L_\nu}{h\nu}~d\nu ~/~ 4\pi~D^2~n_{H}~c,
\end{equation}
   where L$_{\nu}$ is the frequency dependent luminosity of the ionizing 
   continuum, D is the distance between the central source and the
   absorber, 
   n$_{H}$ is the density of atomic hydrogen and
   h$\nu_0$ =13.6 eV. 
Models were generated over the range 10$^{-3.5}$ $\leq$ U$_{abs}$ $\leq$ 10$^{-2}$.
This is quite similar to the range of ionization parameters calculated
for those kinematic components detected in GHRS spectra of NGC 4151 for which both 
C~IV and Mg~II absorption lines could be identified (Kriss 1998).
This range is much lower than the typical values for an X-ray absorber
(U$_{abs}$ $=$ 0.1 - 10, cf. Reynolds \& Fabian 1995).
Integration was truncated at an effective column density (the sum of the 
columns densities of ionized and neutral H),
N$_{eff}$ $=$ 10$^{20}$cm$^{-2}$. This is approximately equal to the
sum of the effective column densities from each kinematic component
detected in NGC 4151 (Kriss 1998).

\subsection{Absorber Model Results}

  The predicted column densities of several ionic species are listed 
in Table 1. Although there are slight differences in the values predicted for
the two different SED's, the results overall show very similar trends, which
is to be expected, since the continua were scaled to produce the
same number of ionizing photons. The C~IV and Mg~II column densities predicted by 
the models with U$_{abs}$ $=$ 10$^{-3}$ are a reasonable match for those calculated for the Seyfert 1.5 galaxy
NGC 4151 (Weymann et al. 1997; Kriss 1998), although they are larger than any {\it single}
component measured by those authors. We should note that our single component
absorber is an idealized model, since a set of absorbers with smaller
effective column densities and different radial velocities, distributed 
along the line-of-sight, would have a similar cumulative effect on the 
ionizing continuum.

 In the most highly ionized cases, where U$_{abs}$ $=$ 10$^{-2}$, the models 
predict C~IV columns
at least an order of magnitude in excess of that observed for NGC 4151
(Weymann et. al
1997). This could be compensated by truncating the models at lower effective
column density, but then the predicted Mg~II columns are too low.
Of course, gas at such a low 
state of ionization could not produce the observed columns of O~VII and O~VIII
seen in many X-ray absorbers (George et al. 1998a). It is possible, then,
that the some of the absorption components are due to traces of 
C~IV and N~V in a X-ray absorber with greater effective column density,
rather than low ionization, optically thin gas (Mathur et al. 1995).

  Although C~IV and Mg~II absorption can coexist in the low-ionization 
models, i.e. with U$_{abs}$ $\leq$ 10$^{-3}$ for both SED's, we did not obtain large columns of
N~V absorption together with Mg~II. NGC 4151 is the only source in which the 
presence of absorption by all three of these
ionic species has been confirmed (cf. Crenshaw et al. 1998). Unfortunately, the spectral
region around N~V $\lambda$1241 has not been observed at sufficiently
high resolution for the identification of individual absorption components 
with those of C~IV $\lambda$1550 and Mg~II $\lambda$2800, although the
low resolution spectra indicate a large column density for N~V ($\sim$ 
10$^{15}$cm$^{-2}$), suggesting a strong high ionization component. We predict that 
the N~V columns from those components that show Mg~II absorption are likely to be below the detection limits
($<$ 10$^{13}$cm$^{-2}$). This prediction will be tested with upcoming
STIS observations of NGC 4151. Thus, is it likely that the N~V absorption,
and possibly some of the C~IV absorption,
occurs in the X-ray absorber or in more highly ionized, optically thin UV absorbing gas,
both of which will be transparent to the ionizing continuum below a few
hundred eV. 

   These models span a range of UV absorber properties, which also show
a range of effects on the EUV continuum. The results are also listed
in Table 1, and displayed graphically in Figures 1 and 2, for the power-law
and BBB models, respectively. Comparing the
ratio of incident ionizing radiation to that which escapes the back-end of 
the absorber at different energies (f$_{E}$ in Table 1), we 
see that, as expected, first the absorption builds above the He~II Lyman limit, 
then near the hydrogen
Lyman limit, and more gradually at energies above 100 eV. In Figures 1 and 2,
the absorption edges of He~I (24.6 eV), O~II (35.1 eV), and C~III (47.4 eV)  
are also clearly visible in the filtered spectrum
from the U$_{abs}$ $=$ 10$^{-3}$ model. 
There are two important effects resulting from the absorption of the ionizing continuum.
First, there is a decrease in the fraction of ionizing radiation reaching
the NLR. To get a quantitative measure of this effect, consider a typical 
narrow line cloud, of density n$_{H}$ $=$ 10$^{5}$ cm$^{-3}$, with an
ionization parameter U$_{nlr}$ $=$ 10$^{-2.5}$ (although one could use any set
of conditions for this comparison). Table 1 also shows the ionization parameter
for a cloud of the same density, at the same distance from the ionizing source, 
if it were screened from the source by the UV absorber. For the most 
extreme case, U$_{nlr}$ can be reduced by approximately two orders of magnitude.
Therefore, for objects with the same intrinsic SED, the presence of a low 
ionization UV absorber with large covering factor can cause dramatic 
differences in the conditions in the NLR, and the resulting emission line
spectrum. The effect will be particularly pronounced for line ratios that
are good ionization parameter diagnostics, such as 
[O~III] $\lambda$5007/[O~II] $\lambda$3727 (cf. Ferland \& Netzer 1983).

   The second spectral property that these absorber models predict
is the absorption of the soft X-ray continuum below 100 eV, primarily
by He~II. At low spectral resolution (i.e. $\sim$ 40\%), such as that provided by 
the {\it ROSAT} Position Sensitive Proportional Counter (PSPC), this
absorption would be manifested by an apparent flattening of the observed continuum. In Figures 3 and 4 we
compare the incident and transmitted continua from 100 eV to 5 keV
for the four UV absorber models, as indicated by the value of the ionization
parameter. As the He~II absorption edge builds, the low energy end of this
band of the ionizing continuum becomes increasingly suppressed. 
Neutral hydrogen is an important source of opacity for the 
lowest ionization models. We can obtain a quantitative measure 
of the continuum flattening by calculating the spectral index from 0.1 to
2.4 keV (to match the {\it ROSAT}/PSPC spectral range), 
$\alpha_{soft-Xray}$, for a linear (i.e. power law) fit to 
log(F$_{\nu}$) vs. log($\nu$). For 
the power law 
models, the unattenuated continuum can be fit with $\alpha_{soft-Xray}$ $\sim$ 
$-$1.38. For comparison, the transmitted continuum for the UV absorber with
U$_{abs}$ $=$ 10$^{-3}$
has an $\alpha_{soft-Xray}$ $\sim$ $-$0.86. If the covering factor for the
UV absorber is 90\%, rather than unity, $\alpha_{soft-Xray}$ becomes
$-$0.97 for the U$_{abs}$ $=$ 10$^{-3}$ model. For the BBB models, the results are similar 
($\alpha_{soft-Xray}$ $\sim$ $-$2.07, unattenuated, and $-$1.58, $-$1.68, for
the U$_{abs}$ $=$ 10$^{-3}$ model, 100\% and 90\% covering, respectively). It is worthwhile to note
that there would be no change to the index calculated from a power-law
fit between the non-ionizing UV (energies $<$ 13.6 eV) and the X-ray (energies $>$ 2 keV), $
\alpha_{UV-2keV}$ (see Section 3.2), if
the absorber were {\it dust-free}, such as those modeled in this paper.

  To summarize, the presence of a UV absorber with a large covering factor
along the line of sight will modify the soft X-ray band of the ionizing continuum.
The lower the ionization state of the absorber, the more pronounced the
effect, which we have illustrated by holding the N$_{eff}$ fixed. 
Also, if the covering factor of the absorber is large along the
line of sight to the NLR, the fraction of ionizing photons reaching the
NLR is inversely proportional to the ionization parameter of the absorber. 
This would have a profound effect on the narrow emission line ratios.
However, when viewed in the context of a sample of objects, the
effects of the modified SED may be diluted by 
by variations in the physical conditions in the NLR gas 
among Seyfert galaxies. A third effect of the UV absorber is to change 
the SED below 100 eV, as indicated by Figures 1 and 2 and the values of f$_{E}$ in Table 1. To examine 
the results of this effect on the conditions in the NLR, we have generated a 
second set of 
photoionization models. The predictions of these models are discussed in the
following section.

\subsection{Narrow Line Models}

   It is well known that the conditions in the Narrow Line Region (NLR) of 
Seyfert galaxies are affected by processes other than photoionization
by the central source, such as starbursts (cf. Heckman et al 1997),
collisional effects such as shocks (cf. Allen, Dopita \& Tsvetanov  1998) 
and heating by cosmic rays (Ferland \& Mushotzky 1984). Also, line ratios
can be affected by the relative contribution from matter-bounded gas 
(Binette, Wilson \& Storchi-Bergmann 1996). Nevertheless, it is likely that 
photoionization is the dominant mechanism for ionization and heating in the 
NLR, as Ferland has argued
(cf. Ferland \& Netzer 1983). This is supported by detailed models of 
individual objects (cf.
Kraemer et al. 1998a), which also show that the composite emission-line 
spectrum appears to be dominated by radiation-bounded gas. Furthermore, the 
NLR models that we present are intended to examine the effects of modification
of the SED by an intervening absorber, and therefore we are foremost concerned
with photoionized gas. The interpretation of observational evidence
for this effect, however, requires us to address some of these concerns, which
we will do in the Discussion section. 

   The predicted emission line ratios from the photoionization models of 
typical NLR clouds are listed in Tables 2 and 3 (for comparison to
the NLR spectra of Seyfert 1.5s, see Cohen (1983)). For each intrinsic
SED, NLR models of gas of density n$_{H}$ $=$ 10$^{5}$cm$^{-3}$ and 
solar abundances (see section 2.1) were generated for the case of an 
unattenuated source, and both 100\%  and 90\% covering by each of the four 
different UV absorber models. The models were truncated at an N$_{eff}$
$=$ 10$^{21}$cm$^{-2}$, which we have found to be a reasonable
average for a mix of radiation- and matter-bounded gas in the NLR of
Seyferts (cf. Kraemer et al. 1998a). Since we are primarily interested in the gross effect of the SED on the physical
conditions in the NLR, only a subset of the emission-line ratios predicted 
by the models are listed. In order to make the comparison of the NLR model 
predictions simpler, we held the ionization parameter for the NLR cloud fixed at 
U$_{nlr}$ $=$ 10$^{-2.5}$. Holding U fixed required scaling up the flux as the UV absorbers
screen out more of the ionizing radiation. This was done by decreasing the
distance between the NLR cloud and the ionizing source. The absorption of the 
continuum was so extreme for model 4 that the scaling became unrealistic, 
effectively placing the NLR cloud {\it closer} to the ionizing source than the 
absorber. Each model predicted the same average H$\beta$ 
emissivity (with the exception of the power-law SED Model 4, which had
a slightly lower emissivity due to the higher fractional ionization of 
hydrogen).

   The narrow He II $\lambda$4686/H$\beta$ ratio is quite sensitive 
to the shape of the ionizing continuum, since neutral hydrogen is the dominant
source of opacity near 13.6 eV, while ionized helium is above 54.4 eV. 
Comparing the model predictions for this ratio provides us with the
most direct way to see the effect of the different SED's. He~II $\lambda$4686/H$\beta$
for the unattentuated case is 0.21 and 0.25 for the power-law and BBB
models, respectively, which is approximately the value expected from
a photon counting calculation (cf. Kraemer \& Harrington 1986). As the 
ionization parameter of the absorber decreases, we start to see the
expected effect on this line ratio, with the value dropping to 0.07 for Model 2
for the case of full source coverage by the UV absorber. Then, as the 
attenuation of the continuum becomes more extreme, the 
He~II $\lambda$4686/H$\beta$ ratio begins to increase up to three times
the value predicted by the unattentuated model for the most absorbed
continuum. Although this is partially an artifact of the scaling used
to fix the ionization parameter of the NLR models, there is also a reflection
of the physical conditions in X-ray ionized gas. In the most extreme case,
i.e. Model 4, the modified ionizing continua includes a significant remnant
of photons near 100 eV, where He~I and He~II cross-sections dominate, but 
effectively no photons at the Lyman limit, resulting in an unusual mix of low 
and high ionization species. Finally, as expected, the inclusion of a
fraction (10\%) of unattenuated continuum mitigates these effects, although
not to the point where they would be undetectable. 

  Due to the choice of ionization parameter, the high ionization collisional
lines are weak in these models, and we have only listed ratios for
C~IV $\lambda$1550, [Ne~IV] $\lambda$2423 and [Ne~V] $\lambda$3426 in Tables
3 and 4. A similar effect is seen for these lines as for He~II $\lambda$4686,
and the two neon lines track the He~II line most closely, as one
would expect since the ionization potentials of their parent ionic states
are above 54.4 eV. Lines from intermediate ionization states with parent 
ionization potentials well below 54.4 eV, including [O~III] $\lambda$5007, 
[O~III] $\lambda$4363, [O~II] $\lambda$3727, [N~II] $\lambda\lambda$6548,6584,
and [Ne~III] $\lambda$3869, show little variation in relative
strength until the X-ray component of the ionizing continuum dominates,
i.e. in Models 3 and 4. The low ionization lines such as [O~I] $\lambda$6300
and Mg~II $\lambda$2800 also show the effects of the increased hardness
of the attenuated continua, although the scaling of model 4 continuum causes
an increase in the ionization fraction that supresses the Mg~II line somewhat.
The oxygen is tightly bound to the neutral fraction of hydrogen by charge
exchange, so the scaling effect is not obvious when looking at oxygen line ratios
relative to H$\beta$.

  The Balmer decrement is strongly affected by collisional excitation in
X-ray heated gas, as noted by Netzer (1982) and Gaskell \& Ferland (1984), and
this is clearly seen in this set of models. Also, Ly$\alpha$/H$\beta$
increases as the ionizing continuum becomes harder, as expected.
It should be noted that very large values of H$\alpha$/H$\beta$ and
Ly$\alpha$/H$\beta$ are predicted for the most extreme cases (Model 4).
Although this fits an extrapolation of the results presented in Gaskell
and Ferland, the code has not been compared in detail with other models
for this extreme case, and, as such, the predictions for these two ratios 
should be taken qualitatively. 

  The differences between the power-law models and corresponding BBB
models are subtle. In the BBB models a larger residual of photons remains
below 100 eV after attentuation by the UV absorber, so the effects on
the NLR gas are less pronounced. Observationally, it would be hard to
distinguish between the two intrinsic SED's on the basis of narrow line
ratios, except, perhaps, from the relative strength of the [O~I] lines,
but these lines depend on the optical depth of the emission line
clouds, which is almost certain to vary among objects.  

\subsection{Summary of Modeling Results}

  From the set of photoionization models described in the previous sections,
we have been able to show that a UV absorber that possesses observable
columns for a range of ionic species, including Mg~II, will attenuate the
ionizing continuum in such a way that the conditions in the NLR will be
affected (assuming a large covering factor). Even when holding the ionization 
parameter fixed for the NLR models, the predicted line ratios show the gross 
effects of the continuum attentuation which should be observable. If we had 
allowed the ionization parameter to vary, as the values for U$_{nlr}$ in Table 1 show, 
the effect on the NLR models would have been even more dramatic.

  The following set of conditions may exist in individual
objects:

 1. If intrinsic UV resonance line absorption is present,
the continuum along the line of sight will be attentuated, near the
He~II Lyman limit for a more highly ionized absorber, then near the 
hydrogen Lyman limit and finally at soft X-ray energies as the absorber
becomes more neutral. Although most of the EUV waveband is unobservable,
{\it ROSAT} and {\it ASCA} observations in the soft X-ray (0.1 to 2 keV)
should show the effects of attenuation by a UV absorber, particularly if
its presence is flagged by Mg~II absorption. 

 2. If the covering factor of the absorber
is large, its effect on the conditions in the NLR gas will be 
observable, particularly in the line ratios most sensitive to the SED such as 
He~II $\lambda$4686/H$\beta$. 

 3. Unless the absorber is dusty or
Compton thick ($>$ 10$^{24}$cm$^{-2}$) there will not be any effect 
on the continuum below the Lyman limit. If the attenuation of the 
X-ray continuum above 2 keV is relatively small, the optical to X-ray 
index, $\alpha$$_{UV-2keV}$, will not be correlated with these other 
observables (although determination of the 2 keV flux depends on
accurate modeling of the low energy X-ray spectrum, and therefore can be
biased by an improper correction for absorption). 

In the following sections we will discuss the observational evidence for 
these conditions.

  These models also predict that the total 
emission from the UV absorbing gas could be comparable to that from the NLR 
gas, assuming that the covering factors are similar. This, of course, depends on the 
density and/or location of the absorber. Nevertheless, the presence
of this emission in the narrow-line spectrum will dilute some of the
effects of the modified ionizing continuum. The predicted emission line 
ratios from the absorber models are listed in Tables 4 and 5. 
In the higher ionization cases, the UV absorber could be a source of emission 
from coronal lines such as [Ne~V] $\lambda$3426 and [Fe~VII] $\lambda$6087.
Interestingly, the conditions for our high ionization parameter models are 
similar to those used in modeling the NLR of NGC 5548 (Kraemer et al. 1998a),
which suggests that some of the C~IV absorption in NGC 5548 might arise
in NLR gas. It is also possible that strong semi-forbidden lines, such as
C~III] $\lambda$1909 and C~II] $\lambda$2326 emission may arise in this
gas, even in the lower ionization cases.

\section{Comparison with Observations}

  In order to compare our model results to the observations most effectively,
we have chosen to concentrate on Seyfert 1.5s (for the sake 
of simplicity, we will refer to all intermediate Seyfert galaxies as type 
1.5s). The presence of X-ray and UV intrinsic absorption has been
detected in both Seyfert 1s and Seyfert 1.5s, and for each subclass the fact 
that we observe emission from the BLR is evidence that we are seeing the 
AGN directly. However, in Seyfert 1.5 spectra it is possible to
separate out the narrow emission-line component much more easily than for
Seyfert 1s. Seyfert 2s are unsuitable since the ionizing source is generally 
obscured (cf. Antonucci 1994) and only a scattered component of the BLR
emission and soft X-ray continuum can be observed, and the latter may be contaminated
by a starburst (Turner et al. 1997). 

  We selected a set of Seyfert 1.5 galaxies for which narrow 
He~II $\lambda$4686/H$\beta$ ratios had been measured. The sample includes the 
13 Seyfert 1.5s studied by Cohen (1983), which is the largest set thereof, 
NGC 4151 (cf. Ferland and Mushotzky 1982), NGC 3516 
(Ulrich \& Pequignot 1980), and LB 1727 (Turner et al. 1998). 
For this set, the values of narrow He~II $\lambda$4686/H$\beta$ range 
from 0.14 to 0.43, with an average close to the values assumed for our
unattenuated NLR models.

\subsection{Narrow Emission Lines}

  A detailed comparison of the observed narrow emission-line ratios of each of the
Seyfert 1.5s in our sample to our model predictions is outside the scope
of this paper (see Kraemer et al. 1998a for an example of this type of
analysis). However, Kraemer et al. (1998b) examined several selected line 
ratios for this set of objects, from which we can make a qualitative 
assessment of the model predictions. First, the relative [O~I] strength 
is weakly anti-correlated with the He II strength, as predicted (with a 
correlation coefficient, r$_{s}$ $=$ $-$0.44 and a probability, P$_{r}$
$=$ 0.12, of exceeded r$_{s}$ in a random sample).
Second,  there is a strong correlation of 
[O~III] $\lambda$5007/[O~II] $\lambda$3727 vs. He~II $\lambda$4686/H$\beta$
(r$_{s}$ $=$ 0.76, P$_{r}$ $=$ 0.002),
similar to a trend that has been noted for the [O~III] $\lambda$5007 vs. 
He~II $\lambda$4686 (Cohen 1983). Although
this could be explained by the contribution from highly ionized, 
matter-bounded gas, as in the case examined by Binette et al. (1996), it
also fits with the predictions for an attenuated ionizing 
continuum. 

\subsection{The Soft X-ray Continuum and He~II $\lambda$4686/H$\beta$}

  In Seyfert galaxies, the hard X-ray spectrum above 2 keV can be
characterized by a fairly flat power law, with a typical index,
$-1.0 \lesssim$ $\alpha$ $\lesssim -0.5$ (e.g., Nandra \& Pounds 1994).
In the soft X-ray band ($\lesssim 1$keV) attenuation by neutral and/or 
ionized material along the line--of--sight becomes important  
(e.g., George et al. 1998a and references therein). 
The flattest X-ray (2-10 keV) spectra have been observed in 
Seyfert 1.5 galaxies (e.g., NGC 4151, Weaver et al. 1994;
NGC 3227, George et al. 1998b) and in most known
cases at least part of the nuclear continuum is heavily absorbed, although 
the connection between absorption and index has not been fully explored. 

   Many Seyfert spectra appear to steepen below $\sim 1$keV 
(e.g. Arnaud et al. 1985, Turner 
\& Pounds 1989, Turner et al. 1998). The luminosities and 
variability seen in the soft X-ray regime suggests 
these ``soft X-ray excesses'' are often dominated by 
an upturn in the underlying continuum, and 
the strength of the soft component for Seyfert 1s varies but the slope appears to be 
well-correlated with the ratio of the UV flux at 1375\AA~ to the flux at 2 
keV (Walter \& Fink 1993). In at least some sources, this soft X-ray component 
may be a power-law which extends to join the UV continuum 
(e.g. Nandra et al. 1995, Zheng et al. 1997, Laor et al. 1997); 
in this case there should
be a correlation between EUV-dependent line ratios and the soft X-ray
index. 

The soft X-ray indices for each of the Seyfert 1.5s in our sample are listed 
in Table 6, and were determined by fits to an absorbed power-law model. Most indices 
were measured from {\it ROSAT}/PSPC data (from 0.1 - 2.4 keV). NGC 3227 and Mrk 6 were derived from 
fits to the {\it ASCA} data (George et al. 1998b, Feldmeier et al. 1998,
respectively) which 
gives a more accurate index as {\it ASCA} allows a better modeling of the 
complex absorption. In the case of 
MCG 8-11-11 the lack of a pointed PSPC observation led us to use the 
0.7 - 10 keV index from the {\it Einstein} Solid State Spectrometer/Monitor
Proportional Counter (Turner et al. 1991). 

   From these results, there appears to be a modest anti-correlation 
between X-ray hardness and the He~II $\lambda$4686/H$\beta$ ratio
(r$_{s}$ $=$ $-$0.47, P$_{r}$ $=$ 0.08), in the sense that 
sources with steep soft X-ray spectra, show relatively strong He~II, as 
shown in Figure 5. However, given the small number of data points,
care should be taken in interpreting this trend.  A more conservative 
interpretation is that there is a zone of exclusion, specifically the 
intersection of flat soft X-ray slope and large
He~II $\lambda$4686/H$\beta$. As we have shown, the presence of 
absorbing material, residing between the X-ray emitting region and the NLR 
could explain this anti-correlation. In fact, the measured soft X-ray
indices for the flatter sources (specifically NGC 4151, NGC 3227, Mrk 6, and 
MCG 8-11-11) are similar to those predicted by our
UV absorber models. If we are truly seeing the results of attentuation,
the implication is that the covering factor is large enough that the NLR is 
exposed to the same continuum as observed in the soft X-ray band.

  In Table 6, we also list the UV to X-ray indices, $\alpha$$_{UV-2keV}$
for those objects for which data were available. As noted, they were either
calculated from the values listed in Walter \& Fink (1993) or from the UV
fluxes at 1460\AA~ in the {\it IUE} low dispersion archive and the 2 keV
luminosity densities determined from {\it Einstein} IPC data by Wilkes et al (1994), assuming an
intrinsic X-ray index of $-$1.0. Although for the most part the $\alpha$$_{UV-2keV}$ and
$\alpha$$_{soft-Xray}$ are similar, there are several cases, once again
those sources with the flattest soft X-ray slopes, where they
disagree. If, as Walter \& Fink's (1993)
assert, the shape of the UV to soft X-ray bump does not vary among
Seyfert 1 galaxies, this result may be interpreted as evidence that 
$\alpha$$_{UV-2keV}$ and $\alpha$$_{soft-Xray}$ match 
when the intrinsic continuum is observed, while mismatches
are the result of attenuation, as our models predict.

\subsection{Intrinsic Absorption}

  As we have seen, there appears to be some correlation between line ratios
and the shape of the soft X-ray continuum that qualitatively fit our
model predictions. A better test of the models is associating
these properties with the intrinsic UV absorption. 

  Among the Seyfert 1.5 galaxies, those with the flattest soft X-ray spectra, 
(specifically NGC 4151, NGC 3227, and MCG 8-11-11) are known to have unusually 
strong intrinsic absorption. {\it IUE} and {\it HST} spectra of NGC 4151 
reveal high column densities for lines 
covering a wide range in ionization (Mg~II to N~V, see Weymann et al. 1997), 
and the presence of absorption by metastable C~III$^{*}$ $\lambda$1175 ((note 
also that Balmer and He~I self absorption are seen in the optical (cf.
Anderson \& Kraft 1969); we have not attempted to model the conditions that would produce these
features). {\it IUE} spectra of NGC 3227 and MCG 8-11-11 also appear to 
show the presence of large absorption columns (Ulrich 1988). Kriss et al. (1997)
report a possible Lyman limit detection in {\it HUT} spectra of
NGC 3227. These objects
all appear to have a mismatch of $\alpha$$_{soft-Xray}$ and 
$\alpha$$_{UV-2keV}$, which, as noted above, is what might be expected
for a continuum modified by a UV absorber. The galaxies with
low ionization absorption can be contrasted with that of objects like
NGC 5548 and NGC 7469, which show absorption by C~IV and N~V, but 
not Si~IV or Mg~II. The $\alpha$$_{soft-Xray}$ and $\alpha$$_{UV-2keV}$
are quite similar in these objects, which is agreement with the results
from our more highly ionized UV absorber models.

   If we include emission line ratios, the overall picture becomes
more complicated. Although most of those with
flatter soft X-ray spectra are weak He~II emitters, NGC 3227 is an exception. This can be
interpreted as indicating that the unattenuated ionizing continuum is 
unusually hard, the NLR gas is matter-bounded, or that the covering factor of the UV/soft X-ray absorber
is low, except along our line of sight. However, our models predict that
in the case of a very low ionization UV absorber the 
narrow He~II $\lambda$/H$\beta$ can be larger in shielded gas than
in gas that is directly exposed to the ionizing source. Also, as the
ionization state of the absorber decreases, the [O~I] $\lambda$6300/H$\beta$ 
ratio will increase. The narrow line spectrum of NGC 3227 shows strong
[O~I] emission (0.8 x H$\beta$, (Cohen 1983)), which indicates that the NLR gas is radiation-bounded
and suggests that the ionization state of the UV absorber may be 
particularly low. It will require {\it HST}/STIS observations to test
this prediction.

  The case of NGC 7469 also merits an explanation. This object, which shows 
the presence of a high ionization state UV absorber (Crenshaw et al. 1998), has both weak He~II and
a steep soft X-ray slope, one that matches the $\alpha$$_{UV-2keV}$ quite
closely. In fact, the He~II is weaker than a simple photon counting
calculation would predict. It is known, however, that a starburst  is
present in NGC 7469 (cf. Wilson et al. 1991). We would suggest that the starburst
is diluting the relative He~II strength, and that He~II $\lambda$4686/H$\beta$
in the AGN ionized gas is probably similar to NGC 5548. 

  As we have noted several times, our analysis of the effect of the
modified ionizing continuum on the NLR gas is based on the assumption that
the covering factor of the low ionization absorber is large (i.e. $>$ 0.5).
Although there are a large fraction of Seyfert 1s
that show intrinsic UV absorption (Crenshaw et al. 1998), which implies
a large covering factor, there is no
a priori reason to assume that the low ionization absorber is associated
with the high ionization UV absorber (or the X-ray absorber). Another
possibility is that the low ionization absorption lines are formed in
an atmosphere above the molecular torus, the existence of which is a
fundamental prediction of the  ``unified'' model for Seyfert galaxies (cf. Antonucci
1994). Kriss et al. (1997) suggest that in several cases (NGC 4151, NGC 3516,
and NGC 3227), our line of sight to the BLR is through this atmosphere,
along the plane of the torus. From this viewing angle, the biconical
distribution of the emission line gas should be quite evident. This
is certainly true for NGC 4151, but less clear for the other cases
(cf. Schmitt \& Kinney 1996). Also, if this model is correct, we should still
see an absorbed soft X-ray continuum when low ionization absorption
is present. The fact that a large fraction
of Seyfert 1.5s have a flat soft X-ray slope appears to support association
of the low ionization gas with the high ionization intrinsic absorption,
a point to which we will return in the Discussion section.

\section{Discussion}
 
  Our basic model predicts that a UV absorber, situated between
the BLR and NLR and with a large covering factor, will have an observable
effect on the soft X-ray continuum and narrow emission-line spectrum.
We have generated a set of model UV absorbers, spanning a range of
ionization parameter. The predicted ionic columns densities, except in the
highest ionization models, are a reasonable match to those observed
in the most heavily obscured Seyfert 1.5 galaxies. We have also demonstrated the
effects of attentuation by this gas on conditions in the NLR, both by 
calculating the ionization parameters for a ``typical'' NLR cloud shielded by these
absorbers and by generating a set of NLR models at fixed ionization
parameter. Also, we have shown that significant soft X-ray attentuation by 
this absorbing layer
can result in an apparent flattening soft x-ray spectrum (when observed
at low spectral resolution), as shown in Figures 3 and 4. 
Finally, there are sources where the full set of predicted characteristics
are present. If this model is correct, we should be able to predict
the presence of one characteristic, such as UV absorption, if the other two
are observed: a low ionization, He~II weak narrow-line spectrum
and a flat soft X-ray spectrum. 

  As noted above, the spectral characteristics of NGC 4151 match the 
model predictions quite well, and MCG 8-11-11 and NGC 3227 also show evidence 
that the same process is at work. 
Specifically,  we predict that more than 1/3 of the Seyfert 1.5s in our sample
(specifically those with $\alpha$$_{soft-Xray}$ $\leq$ $-$1.0; see Table 6)
possess a large column of low ionization UV absorption. If, as Kriss et al. (1997) 
suggest, the low ionization lines form in an atmosphere associated with the
molecular torus, either the atmosphere must extend well above the plane of the
torus or the characteristics of Seyfert 1.5s, e.g. the detection of a 
narrow component in the permitted emission lines, are more easily
observed when the AGN is viewed along this line-of-sight. If the former
were true, the covering factor of the absorber along the line of sight
to the NLR could still be large, and our predictions on the effect in the
NLR gas are relevant. 

\section{Summary}

  We have explored the effect of the gas in which intrinsic UV absorption
  lines arise on the ionizing continuum in Seyfert galaxies. 
  The main results are the following: 

  1. Above 100 eV, the absorber will modify the soft X-ray continuum,
  if the column density is sufficiently large 
  (N$_{eff}$ $\geq$ 10$^{20}$cm$^{-2}$) and the gas is not highly
  ionized (U$_{abs}$ $\leq$10$^{-2.5}$), as is the case for NGC 4151.
  There is observational evidence for this effect, since objects that
  show large columns of low ionization absorbers tend to have
  flatter soft X-ray continuum slopes.

  2. A low ionization absorber will attenuate much of the ionizing
  radiation between 13.6 eV and 100 eV, in particular near the 
  He~II Lyman limit. There is evidence for this in that the
  relative strength of the narrow He~II line is anti-correlated
  with the hardness of the soft X-ray continuum in Seyfert 1.5s.

  3. Since the presence of low ionization UV absorption and a flat soft X-ray
  continuum may be interrelated, one can predict that a Seyfert galaxy will 
  exhibit one of these characteristics if the other is present. If so,
  a large fraction (1/3) of Seyfert 1.5s should show low ionization UV
  absorption lines. Also, if the covering factor of the low ionization
  gas is close to unity, we would expect that these absorbed Seyfert 1.5s
  will possess low ionization narrow emission-line spectra.

\acknowledgments

 S.B.K. thanks Eric Smith for useful discussions.
 S.B.K. and D.M.C. acknowledge support from NASA grant NAG5-4103. T.J.T.
 acknowledges support from UMBC and NASA/LTSA grant NAG5-7835.

\clearpage

\figcaption[power_low.eps]{Spectral Energy Distribution, 8.7 eV -- 100 eV, for the
power law models; the solid line represents the incident continuum and the
dashed line the filtered continuum, for the U $=$ 0.01 and 0.001 models,
N$_{eff}$ $=$ 10$^{20}$cm$^{-2}$. The energies of several absorption edges
are noted.
}\label{fig1} 

\figcaption[bbb_low.eps]{Spectral Energy Distribution, 8.7 eV -- 100 eV, for the
BBB models; the solid line represents the incident continuum and the
dashed line the filtered continuum, for the U $=$ 0.01 and 0.001 models,
N$_{eff}$ $=$ 10$^{20}$cm$^{-2}$.The energies of several absorption edges
are noted.
}\label{fig2} 

\figcaption[sed1.eps]{Spectral Energy Distribution, 100 eV -- 5 keV, for the
power law models; the solid line represents the incident continuum and the
dashed line the filtered continuum, for each of the four models
(N$_{eff}$ $=$ 10$^{20}$cm$^{-2}$ in all cases).
}\label{fig3} 

\figcaption[sed2.eps]{Spectral Energy Distribution, 100 eV -- 5 keV, for the
BBB models; the solid line represents the incident continuum and the
dashed line the filtered continuum, for each of the four models
(N$_{eff}$ $=$ 10$^{20}$cm$^{-2}$ in all cases).
}\label{fig4} 

\figcaption[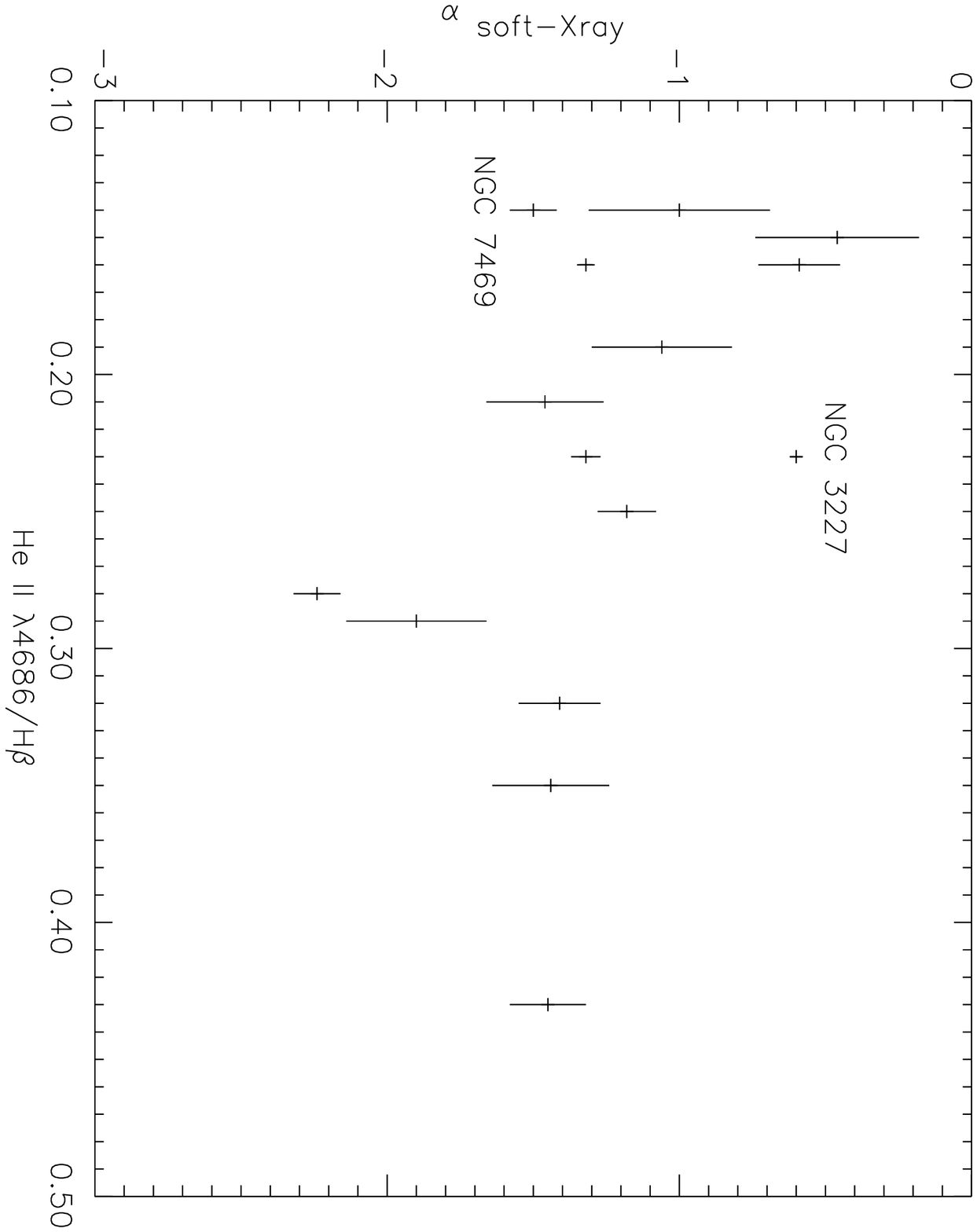]{Soft X-ray Index (from a power-law fit, from 0.1 - 2.4 keV, except as 
noted in text) versus narrow He~II $\lambda$/H$\beta$
in the sample of Seyfert 1.5s; error in He~II/H$\beta$ $\approx$ 25\%.
}\label{fig5}

\clearpage

\begin{deluxetable}{lcccc}
\footnotesize
\tablenum{4}
\tablecolumns{5}
\tablecaption{Emission Line Ratios (relative to to H$\beta$) for UV Absorber 
Models$^{a}$, assuming a Power-law Ionizing Continuum
}
\tablewidth{0pt}
\tablehead{
\colhead{} & \colhead{U$_{abs}$ = 10$^{-2.0}$} & 
\colhead{U$_{abs}$ = 10$^{-2.5}$ } & 
\colhead{U$_{abs}$ = 10$^{-3.0}$ } & 
\colhead{U$_{abs}$ = 10$^{-3.5}$ }
}
\startdata
Ly$\alpha$ $\lambda$1216                     &37.27             &38.14 
              &40.34                         &50.36\\
N V $\lambda$1240                            & 1.21             & 0.07
              & 0.00                         & 0.00\\
Si IV $\lambda$1398                          & 0.85             & 1.68
              & 0.13                         & 0.00\\
O IV] $\lambda$1400                          & 4.18             & 0.95
              & 0.04                         & 0.00\\
N IV] $\lambda$1485                          & 1.75             & 0.41
              & 0.03                         & 0.00\\
C IV $\lambda$1550          	      	     &31.31             & 10.11 
              & 0.67                         & 0.03\\
O III] $\lambda$1664                         & 1.31             & 2.65
              & 1.08                         & 0.21\\
N III] $\lambda$1750                         & 1.47             & 1.75
              & 0.63                         & 0.11\\
C III] $\lambda$1909                         &11.23             & 17.07 
              & 7.28                         & 2.21\\
C II] $\lambda$2326                          & 0.10             & 0.92
              & 3.23                         & 4.91\\
Mg II $\lambda$2800         	      	     & 0.00             & 0.30
              & 3.15                         & 4.91\\
$[$Ne V] $\lambda$3426      	      	     & 2.05             & 0.48 
              & 0.01                         & 0.00\\
$[$Fe VII] $\lambda$3586                     & 0.30             & 0.06
              & 0.00                         & 0.00\\
$[$Fe VII] $\lambda$3760                     & 0.41             & 0.08
              & 0.00                         & 0.00\\
$[$Ne III] $\lambda$3869    	      	     & 0.08             & 0.97 
              & 1.44                         & 1.17\\
$[$O III] $\lambda$4363      	      	     & 0.95             & 2.21 
              & 1.13                         & 0.26\\
He II $\lambda$4686          	      	     & 1.01             & 0.66 
              & 0.23                         & 0.16\\
H$\beta$                 	      	     & 1.00 	           & 1.00 	
    & 1.00          	 & 1.00\\
$[$O III] $\lambda$5007                      & 1.06             & 3.04 
              & 2.11                         & 0.61\\
$[$Fe VII] $\lambda$5721                     & 0.42             & 0.10
              & 0.00                         & 0.00\\
$[$Fe VII] $\lambda$6087                     & 0.63             & 0.14
              & 0.00                         & 0.00\\
$[$O I] $\lambda$6300        	      	     & 0.00             & 0.00 
              & 0.00                         & 0.50\\
H$\alpha$ $\lambda$6563      	      	     & 2.83              & 2.89 
              & 2.96                         & 3.17 \\
\tablenotetext{a}{n$_{H}$=1x10$^{7}$cm$^{-3}$, no dust,
N$_{eff}$ = 10$^{20}$cm$^{-2}$}

\enddata
\end{deluxetable}

\clearpage

\begin{deluxetable}{lcccc}
\footnotesize
\tablenum{5}
\tablecolumns{5}
\tablecaption{Emission Line Ratios (relative to to H$\beta$) from UV Absorber 
Models$^{a}$, assuming a BBB Ionizing Continuum
}
\tablewidth{0pt}
\tablehead{
\colhead{} & \colhead{U$_{abs}$ = 10$^{-2.0}$} & 
\colhead{U$_{abs}$ = 10$^{-2.5}$ } & 
\colhead{U$_{abs}$ = 10$^{-3.0}$ } & 
\colhead{U$_{abs}$ = 10$^{-3.5}$ } 
}
\startdata
Ly$\alpha$ $\lambda$1216                     
                     &37.24             &38.16 
              &41.35                         &52.50\\
N V $\lambda$1240                            
                           & 1.24             & 0.07
              & 0.00                         & 0.00\\
Si IV $\lambda$1398                          
               & 1.65             & 1.99
              & 0.17                         & 0.00\\
O IV] $\lambda$1400                          
                          & 3.85             & 1.11
              & 0.06                         & 0.00\\
N IV] $\lambda$1485                          
                          & 1.74             & 0.50
              & 0.04                         & 0.00\\
C IV $\lambda$1550          	      	     
	      	     &29.45             & 12.34 
              & 1.06                         & 0.05\\
O III] $\lambda$1664                         
                        & 0.89             & 2.22
              & 1.25                         & 0.27\\
N III] $\lambda$1750                         
              & 1.03             & 1.50
              & 0.72                         & 0.14\\
C III] $\lambda$1909                         
                        & 7.38             & 14.82 
              & 8.31                         & 2.71\\
C II] $\lambda$2326                          
                         & 0.07             & 0.67
              & 2.82                         & 4.86\\
Mg II $\lambda$2800         	      	     
	      	     & 0.00             & 0.28
              & 3.06                         & 4.92\\
$[$Ne V] $\lambda$3426      	      	     
    	      	     & 2.09             & 0.43 
              & 0.01                         & 0.00\\
$[$Fe VII] $\lambda$3586                     
                    & 0.29             & 0.05
              & 0.00                         & 0.00\\
$[$Fe VII] $\lambda$3760                     
                     & 0.40             & 0.06
              & 0.00                         & 0.00\\
$[$Ne III] $\lambda$3869    	      	     
    	      	     & 0.08             & 0.84 
              & 1.51                         & 1.26\\
$[$O III] $\lambda$4363      	      	     
     	      	     & 0.65             & 1.87 
              & 1.28                         & 0.32\\
He II $\lambda$4686          	      	     
	      	     & 1.03             & 0.78 
              & 0.29                         & 0.20\\
H$\beta$                 	      	     & 1.00 	           & 1.00 	
    & 1.00          	 & 1.00\\
$[$O III] $\lambda$5007                      
                     & 0.71             & 2.61 
              & 2.33                         & 0.61\\
$[$Fe VII] $\lambda$5721                     
                     & 0.41             & 0.07
              & 0.00                         & 0.00\\
$[$Fe VII] $\lambda$6087                     
                     & 0.62             & 0.10
              & 0.00                         & 0.00\\
$[$O I] $\lambda$6300        	      	     
        	      	     & 0.00             & 0.00 
              & 0.00                         & 0.45\\
H$\alpha$ $\lambda$6563      	      	     
     	      	     & 2.83              & 2.89 
              & 3.00                         & 3.23 \\
\tablenotetext{a}{n$_{H}$=1x10$^{7}$cm$^{-3}$, no dust,
N$_{eff}$ = 10$^{20}$cm$^{-2}$}

\enddata
\end{deluxetable}

\clearpage

\begin{deluxetable}{lccc}
\scriptsize
\tablenum{6}
\tablecolumns{4}
\tablecaption{Seyfert 1.5s: Soft X-ray Indices}
\tablewidth{0pt}
\tablehead{
\colhead{name} & \colhead{narrow He~II $\lambda$4686/H$\beta$$^{a}$} & 
\colhead{$\alpha$$_{soft-Xray}$$^{b}$} &
\colhead{$\alpha$$_{UV-2keV}$}
}
\startdata
NGC 3227  	& 0.23 $\pm$0.07 & $-$0.60$^{c}$ $\pm$0.02 &$-$1.21$^{k}$\\
NGC 3516        & 0.35$\pm$0.10  & $-$1.44$^{d}$ $\pm$0.20 &$-$1.58$^{k}$\\
NGC 4151        & 0.19$\pm$0.06  & $-$1.06$^{d}$ $\pm$0.24 &$-$1.64$^{k}$\\  
NGC 5548        & 0.23$\pm$0.05  & $-$1.32$^{d}$ $\pm$0.05 &$-$1.27$^{l}$\\
NGC 7469        & 0.14$\pm$0.04  & $-$1.50$^{f}$ $\pm$0.08 &$-$1.52$^{l}$\\
Mrk 6           & 0.15$\pm$0.04 & $-$0.46$^{e}$ $\pm$0.87 & \\
Mrk 79          & 0.25$\pm$0.07 & $-$1.18$^{d}$ $\pm$0.10 &$-$1.30$^{l}$\\
Mrk 279         & 0.16$\pm$0.05 & $-$1.32$^{g}$ $\pm$0.03 &$-$1.36$^{l}$\\
Mrk 506         & 0.21$\pm$0.06 & $-$1.46$^{h}$ $\pm$0.20 &$-$1.32$^{l}$\\
Mrk 704         & 0.43$\pm$0.12 & $-$1.45$^{h}$ $\pm$0.13 &$-$1.36$^{l}$\\
Mrk 817         & 0.29$\pm$0.08 & $-$1.90$^{d}$ $\pm$0.24 &\\
Mrk 841         & 0.32$\pm$0.09 & $-$1.41$^{i}$ $\pm$0.14 &$-$1.70$^{l}$\\
Mrk 926         & 0.14$\pm$0.04 & $-$1.00$^{h}$ $\pm$0.31 &$-$1.35$^{l}$\\
MCG 8-11-11     & 0.16$\pm$0.05 & $-$0.59$^{j}$ $\pm$0.14 &$-$0.95$^{l}$\\        
LB 1727         & 0.28$\pm$0.06 &$-$2.2$^{f}$$\pm$0.08 &  \\
\tablenotetext{a}{errors as quoted by Cohen (1983).}
\tablenotetext{b}{see text (Section 3.2) for the description of the
derivation of $\alpha$$_{soft-Xray}$.}
\tablenotetext{c}{{\it ASCA} (George et al. 1998b).}
\tablenotetext{d}{{\it ROSAT}/PSPC (Rush et al. 1996, see text).}
\tablenotetext{e}{{\it ASCA} (Feldmeier et al. 1998).}
\tablenotetext{f}{{\it ROSAT}/PSPC (Turner, George \& Mushotzky 1993).}
\tablenotetext{g}{{\it ROSAT}/PSPC (this paper).}
\tablenotetext{h}{{\it ROSAT}/PSPC (Walter \& Fink 1993).}
\tablenotetext{i}{{\it ROSAT}/PSPC (Nandra et al. 1995).}
\tablenotetext{j}{SSS/MPC (Turner et al. 1991).}
\tablenotetext{k}{fit between 1460\AA~, from {\it IUE} and 2 keV, from
{\it Einstein} IPC, assuming an intrinsic $\alpha$ $=$ $-$1.0 in the
X-ray (Wilkes 
et al 1994).}
\tablenotetext{l}{fit between 1375\AA~ and 2 keV from 
values in Walter \& Fink (1993) (see references therein).}

\enddata
\end{deluxetable}

\clearpage

\clearpage
\plotone{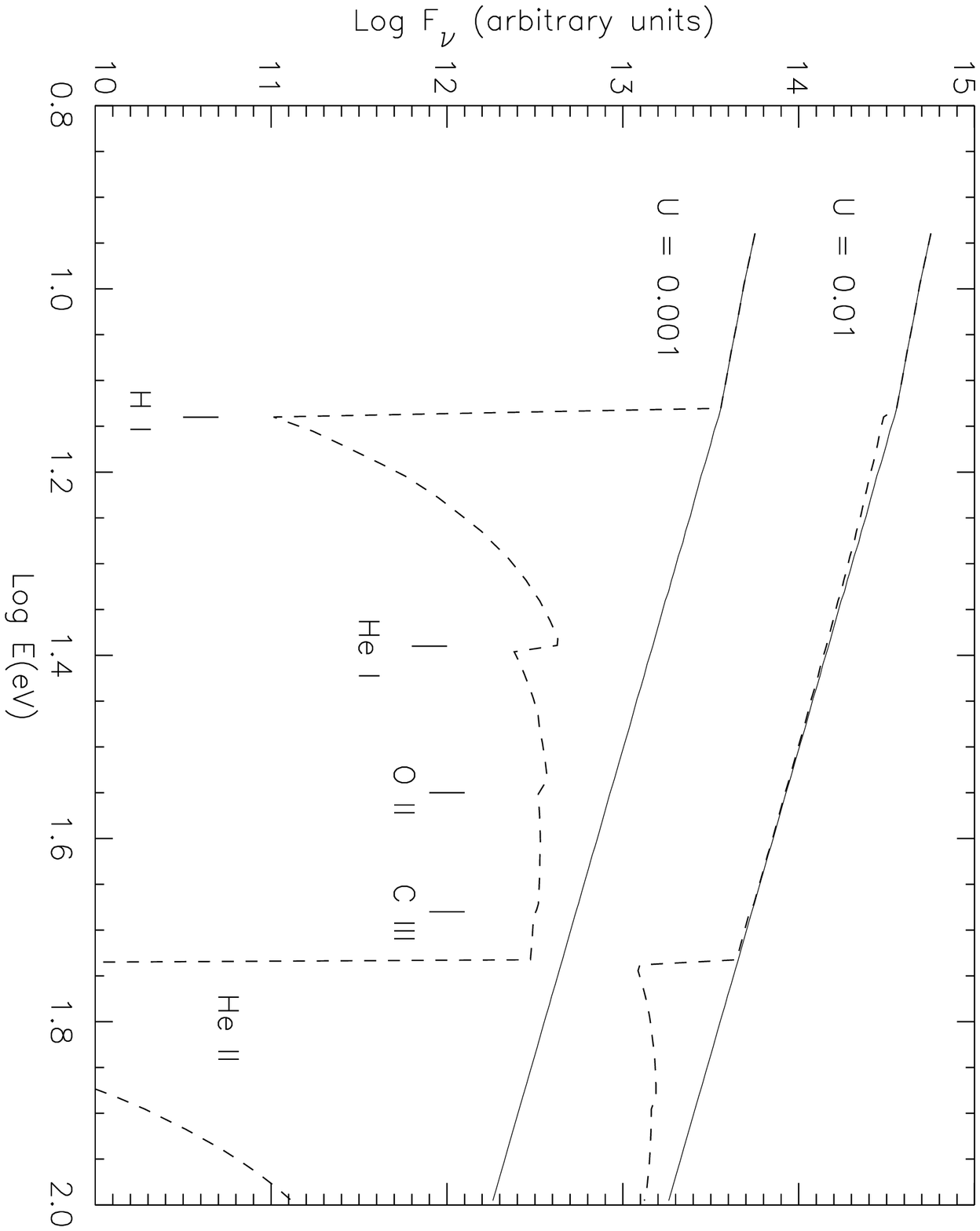}

\clearpage
\plotone{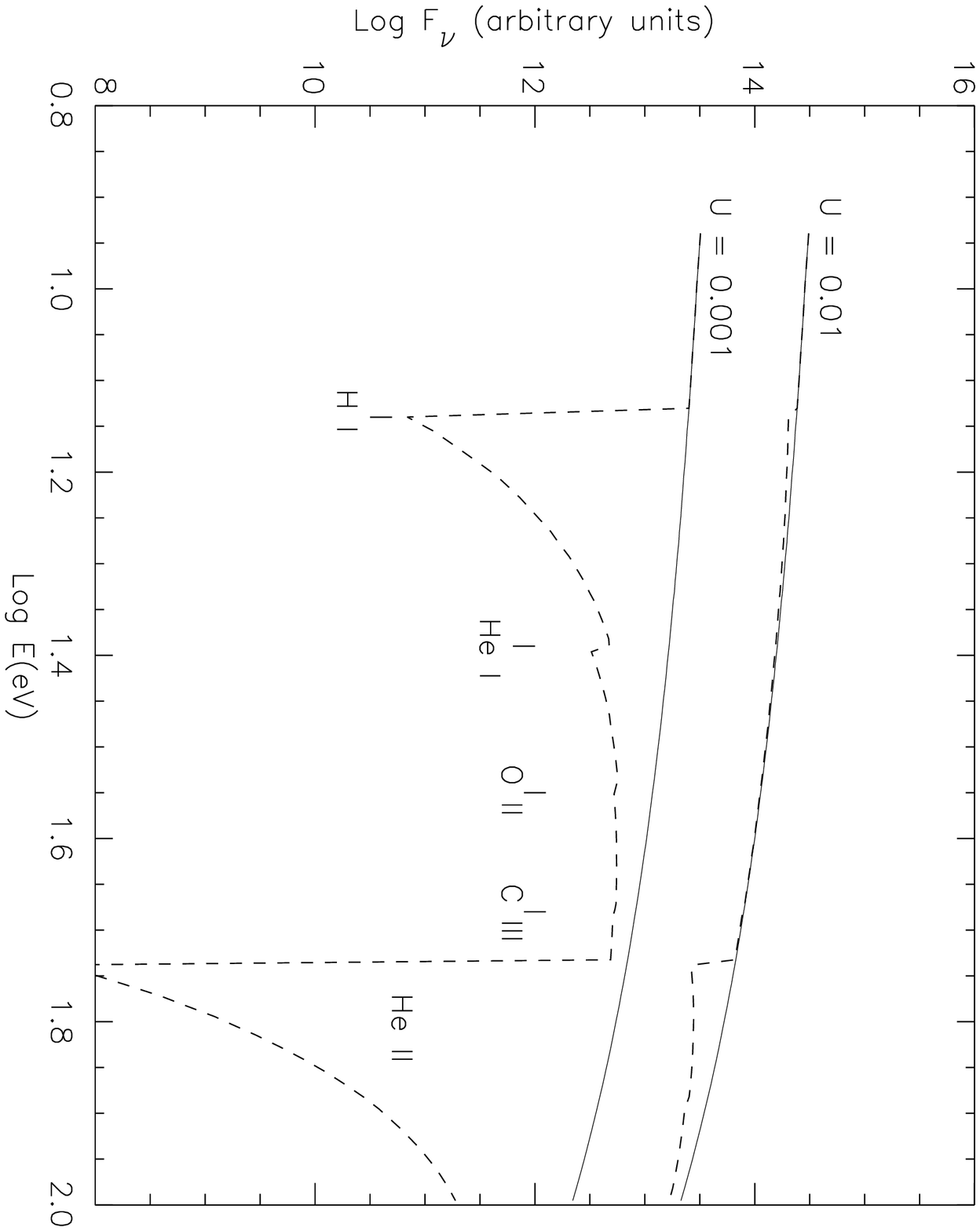}

\clearpage
\plotone{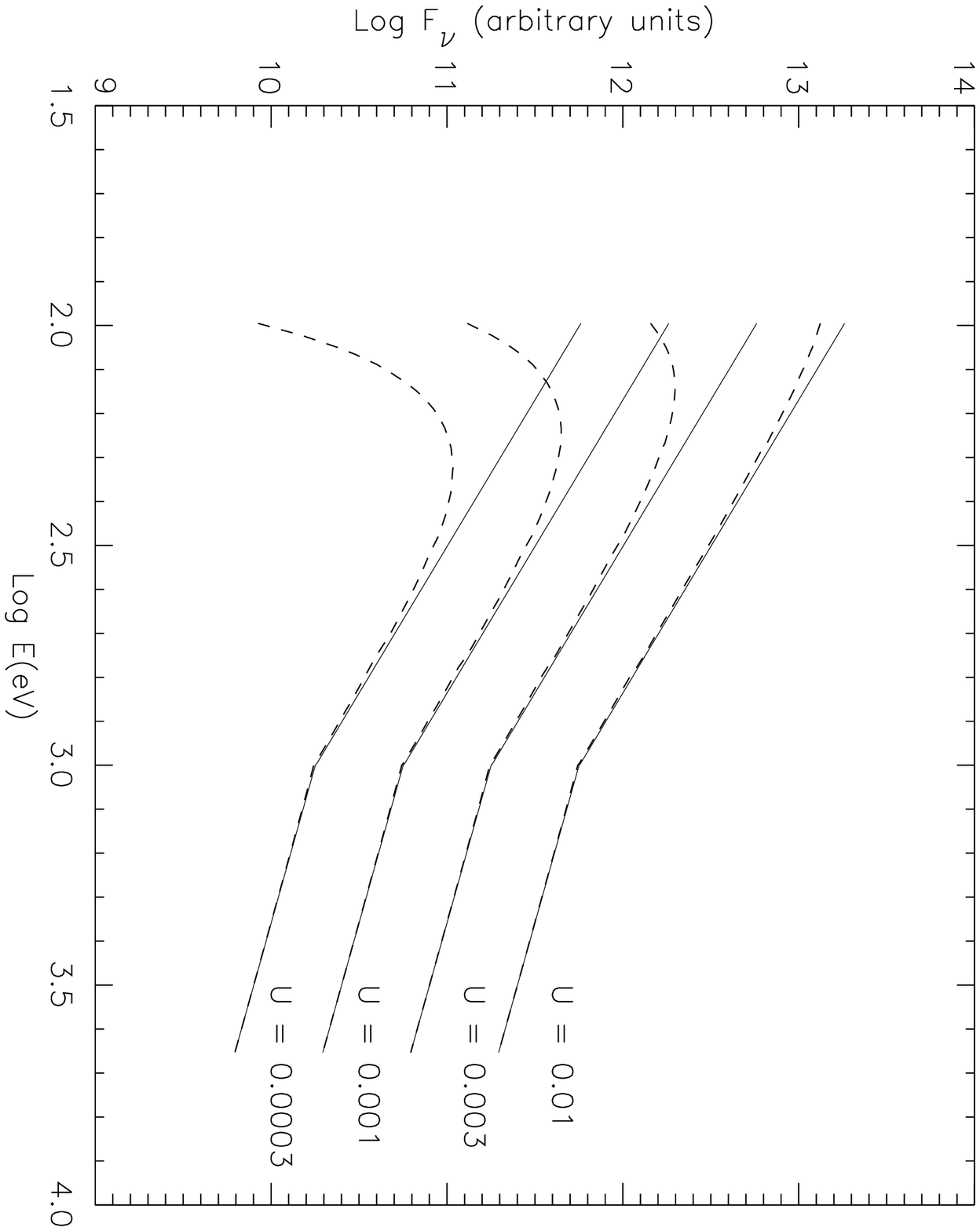}

\clearpage
\plotone{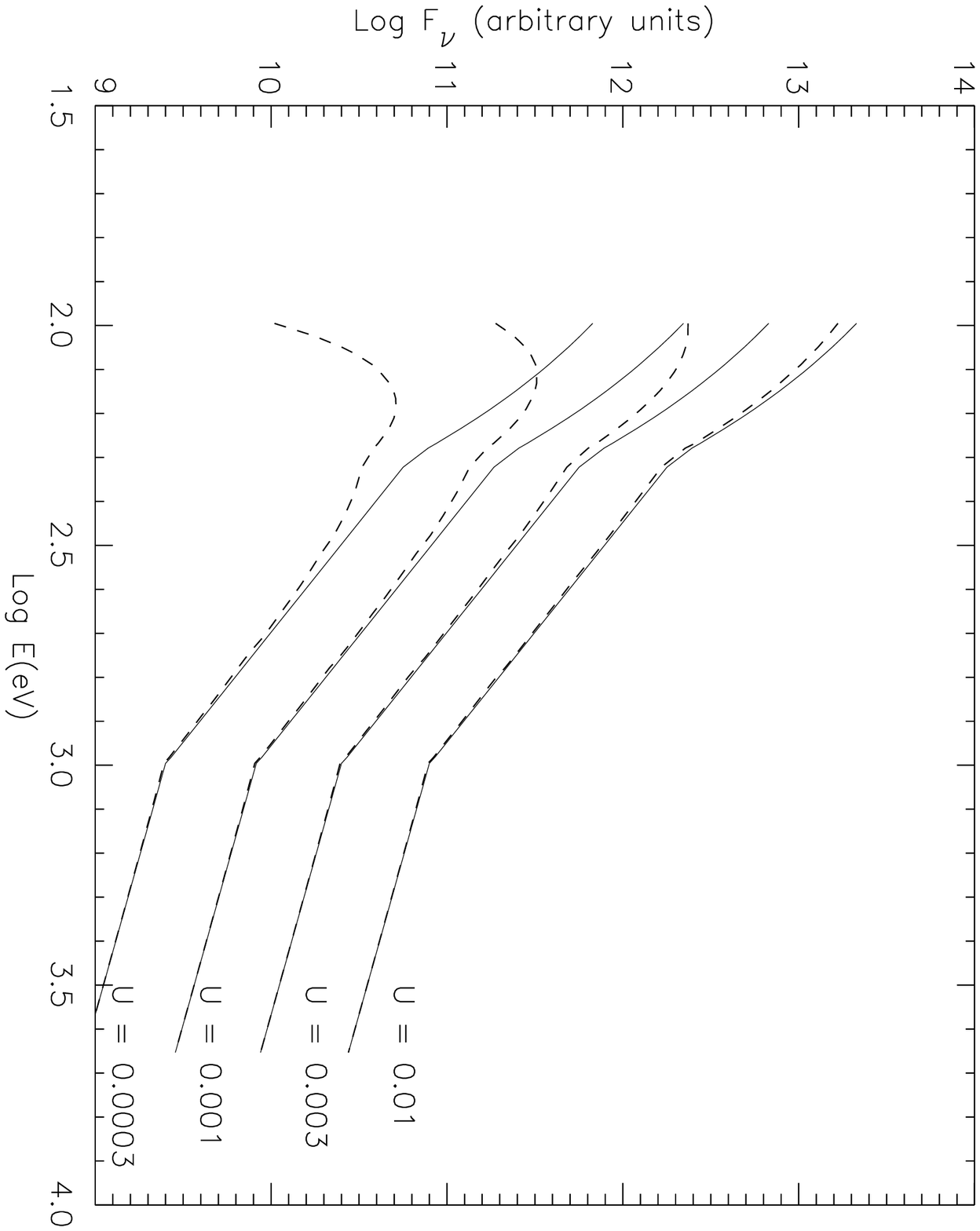}

\clearpage
\plotone{fig5.eps}



\end{document}